\documentclass[twocolumn,aps,pra,superscriptaddress,showpacs]{revtex4-2}
\usepackage{epsfig}
\usepackage[colorlinks=true,citecolor=blue,urlcolor=blue,linkcolor=blue]{hyperref}
\usepackage{graphicx}
\usepackage{multirow}
\usepackage[table,xcdraw]{xcolor}
\usepackage{natbib}
\usepackage{chemfig}
\usepackage[version=4]{mhchem}
\usepackage[normalem]{ulem}
\usepackage{xcolor} 
\usepackage{xspace}
\definecolor{darkgreen}{rgb}{0.0, 0.6, 0.0}

\DeclareMathAlphabet{\mathbsf}{OT1}{cmss}{bx}{n}

\newcommand{\phenyl}{\raisebox{0.21ex}{-}\raisebox{0.75ex}{\scalebox{0.175}{\chemfig[height=1.7ex]{[:-30]**6(------)}}}}
\newcommand{\phenylX}[1]{\raisebox{0.21ex}{-}\raisebox{0.75ex}{\scalebox{0.175}{\chemfig[height=1.7ex]{[:-30]**6(------)}}}\raisebox{0.21ex}{-}#1}
\newcommand{\twonaphthyl}{\raisebox{0.21ex}{-}\raisebox{0.75ex}{\scalebox{0.175}{\chemfig[height=1.7ex]{[:-30]**6(=-**6(-=-=-)=-=-)}}}}
\newcommand{\onenaphthyl}{\raisebox{0.21ex}{-}\raisebox{0.75ex}{\scalebox{0.175}{\chemfig[height=1.7ex]{[:-30]**6(=**6(-=-=-)-=-=-)}}}}
\newcommand{\cm}{cm$^{-1}$\xspace}
\newcommand{\A}{$\widetilde A$\xspace}
\newcommand{\B}{$\widetilde B$\xspace}
\newcommand{\C}{$\widetilde C$\xspace}
\newcommand{\X}{$\widetilde X$\xspace}

\begin{document}

\title{Pathway Towards Optical Cycling and Laser Cooling of Functionalized Arenes}
\author{Debayan Mitra}
\altaffiliation{Present address: Department of Physics, Columbia University, New York}
\affiliation{Harvard-MIT Center for Ultracold Atoms, Cambridge, MA 02138, USA}
\affiliation{Department of Physics, Harvard University, Cambridge, MA 02138, USA}

\author{Zack D. Lasner}
\affiliation{Harvard-MIT Center for Ultracold Atoms, Cambridge, MA 02138, USA}
\affiliation{Department of Physics, Harvard University, Cambridge, MA 02138, USA}

\author{Guo-Zhu Zhu}
\affiliation{Department of Physics and Astronomy, University of California, Los Angeles, California 90095, USA}
\affiliation{Center for Quantum Science and Engineering, University of California, Los Angeles, California 90095, USA}
\affiliation{Challenge Institute for Quantum Computation, University of California, Los Angeles, California 90095, USA}

\author{Claire E. Dickerson}
\affiliation{Department of Chemistry and Biochemistry, University of California, Los Angeles, California 90095, USA}

\author{Benjamin L. Augenbraun}
\affiliation{Harvard-MIT Center for Ultracold Atoms, Cambridge, MA 02138, USA}
\affiliation{Department of Physics, Harvard University, Cambridge, MA 02138, USA}

\author{Austin D. Bailey}
\affiliation{Department of Chemistry and Biochemistry, University of California, Los Angeles, California 90095, USA}

\author{Anastassia N. Alexandrova}
\affiliation{Center for Quantum Science and Engineering, University of California, Los Angeles, California 90095, USA}
\affiliation{Department of Chemistry and Biochemistry, University of California, Los Angeles, California 90095, USA}

\author{Wesley C. Campbell}
\affiliation{Department of Physics and Astronomy, University of California, Los Angeles, California 90095, USA}
\affiliation{Center for Quantum Science and Engineering, University of California, Los Angeles, California 90095, USA}
\affiliation{Challenge Institute for Quantum Computation, University of California, Los Angeles, California 90095, USA}

\author{Justin R. Caram}
\affiliation{Center for Quantum Science and Engineering, University of California, Los Angeles, California 90095, USA}
\affiliation{Department of Chemistry and Biochemistry, University of California, Los Angeles, California 90095, USA}

\author{Eric R. Hudson}
\affiliation{Department of Physics and Astronomy, University of California, Los Angeles, California 90095, USA}
\affiliation{Center for Quantum Science and Engineering, University of California, Los Angeles, California 90095, USA}
\affiliation{Challenge Institute for Quantum Computation, University of California, Los Angeles, California 90095, USA}

\author{John M. Doyle}
\affiliation{Harvard-MIT Center for Ultracold Atoms, Cambridge, MA 02138, USA}
\affiliation{Department of Physics, Harvard University, Cambridge, MA 02138, USA}
\date{\today}

\begin{abstract}
Rapid and repeated photon cycling has enabled precision metrology and the development of quantum information systems using atoms and simple molecules. Extending optical cycling to structurally complex molecules would provide new capabilities in these areas, as well as in ultracold chemistry. Increased molecular complexity, however, makes realizing closed optical transitions more difficult. Building on already established strong optical cycling of diatomic, linear triatomic, and symmetric top molecules, recent work has pointed the way to cycling of larger molecules, including phenoxides. The paradigm for these systems is an optical cycling center bonded to a molecular ligand. Theory has suggested that cycling may be extended to even larger ligands, like naphthalene, pyrene and coronene. Here, we study optical excitation and fluorescent vibrational branching of CaO\twonaphthyl, SrO\twonaphthyl~and CaO\onenaphthyl{} and find only weak decay to excited vibrational states, indicating a promising path to full quantum control and laser cooling of large arene-based molecules.
\end{abstract}

\maketitle
The repeated absorption and emission of photons by atoms is the fundamental process that underlies many advances in atomic, molecular, and optical science (AMO). This optical cycling enables the technique of laser cooling, as used for the deceleration of atoms emanating from a hot oven \cite{Phillips1982Laser}, and magneto-optical trapping \cite{Raab1987mot}. It is further employed for quantum state preparation and readout \cite{Bernien2017,Monroe2021,Morgado2021}, which enable ultra-precise clocks \cite{Young2020clock,Ohmae2021clock,Brewer2019clock}, atom-based quantum simulators and computers \cite{Chiu2019string,Semeghini2021simulator,Grzesiak2020ionq}, and studies of single-quantum-state-controlled ultracold chemistry~\cite{Balakrishnan2016chemistry,Liu2021chemistry,Heazlewood2021chemistry}.

Extending optical cycling to a wide variety of molecules is expected to bring a wealth of new science arising from their rotational and vibrational states. However, the complexity of molecules, as compared to atoms, brings new challenges for optical cycling. In particular, a molecule that is optically driven by a narrow band laser to an excited electronic state can decay to excited rotational and vibrational states in the electronic ground state. Thus, the molecule can end up in ``dark" states that are no longer excited by the laser. Despite the difficulties of these dark state processes (i.e. ``leakage" to excited rovibrational states), optical cycling and laser cooling of molecules has been accomplished by making use of symmetry driven selection rules and multi-wavelength laser excitation to ``repump" rotational and vibrational dark states~\cite{mccarron2018laser}. The diatomic molecules SrF~\cite{shuman2010laser}, CaF~\cite{truppe2017molecules,anderegg2017radio}, and YO~\cite{collopy20183d} were the first molecules to be directly laser cooled using this approach.

\begin{figure*}[ht!]
    \centering
    \includegraphics[scale=1]{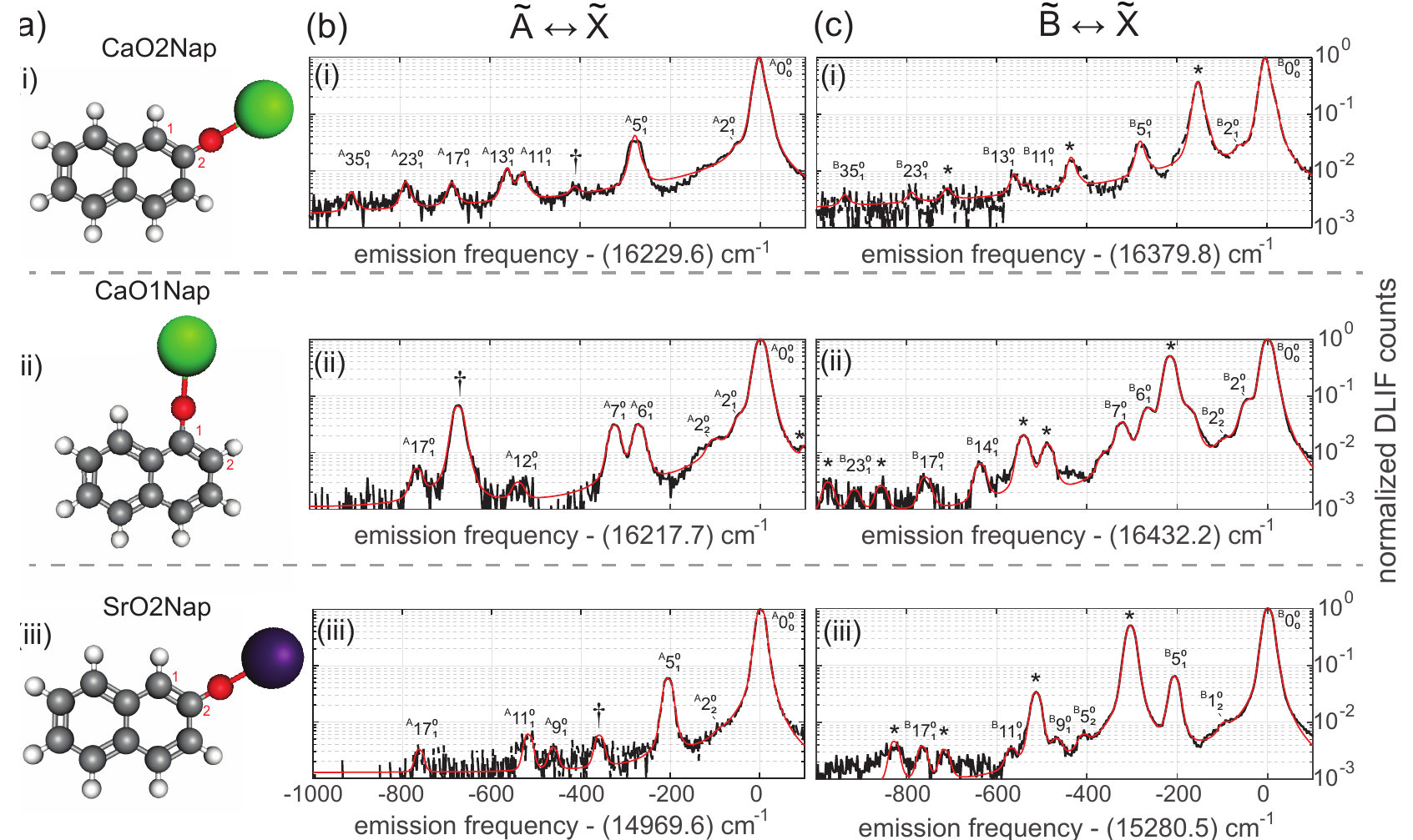}
    \caption{DLIF spectra of molecules studied in this work. (a(i)-(iii)) are ball-and-stick representations of CaO2Nap, CaO1Nap and SrO2Nap, respectively. White spheres denote hydrogen, gray denote carbon, red denote oxygen, green denote calcium and purple denote strontium atoms. Also labelled are the only two distinguishable carbon bonding sites in naphthalene (1 and 2). (b(i)-(iii)) are  DLIF spectra after exciting the \A$\leftarrow$ \X transition for the molecular species (a(i)-(iii)) normalized to the peak height at 0 \cm. Black lines are experimental data that result from averaging $\sim$ 5000 repetitions. (c(i)-(iii)) are \B$\leftarrow$ \X excitations. All attributed peaks have been labelled as $^S$v$^i_f$, where the first superscript defines the excited state (\A or \B), the number v denotes the v$^\text{th}$ vibrational mode, the second superscript $i$ defines the vibrational quantum number in the excited state (always the vibrational ground state (v = 0) in the \A or \B state), and the subscript $f$ is the vibrational quantum number in the ground \X state that the molecule decays to. Red lines represent fits to the DLIF spectra used to extract the VBRs for each decay pathway. Peaks marked with $\dagger$ represent contaminant peaks that cannot be ascribed to theoretically estimated harmonic modes or the overtones of the molecule of interest. The peaks marked with $*$ are due to collision-induced transitions between the \B and the \A states. The peaks marked $*$ and $\dagger$ have been excluded from VBR estimation.}
    \label{fig:fig1}
\end{figure*}

During the development of laser cooling of diatomic molecules, experimental work toward laser cooling of polyatomic molecules began with SrOH \cite{kozyryev2015collisional}. Soon after, it was proposed that low leakage to excited vibrational states (i.e. diagonal Franck-Condon factors) could be obtained in a class of alkaline earth (I) ``metal-oxide-radical" polyatomic molecules, of which SrOH is an example~\cite{Kozyryev2016polyatomic}. The general concept of the use of non-bonding orbitals for quasi-closed transitions in polyatomic molecules was also proposed~\cite{isaev2016polyatomic}. Since then, laser cooling has been successfully demonstrated in the linear triatomic molecules SrOH \cite{kozyryev2017sisyphus}, CaOH \cite{baum20201d, vilas2021CaOHMOT}, and YbOH \cite{Augenbraun2020sisyphus}, and the symmetric top molecule CaOCH$_3$ \cite{mitra2020direct}. Recently, CaOH was laser cooled to temperatures $<1$~mK and confined in a MOT by repumping only nine vibrational loss channels (scattering $\sim10^4$ photons) \cite{vilas2021CaOHMOT,Baum2021Establishing}.

Two features of these molecular systems allow for successful optical cycling.  First, alkaline-earth(-like) metals (such as Ca, Sr, or Yb) act as an optical cycling center (OCC) that is bonded to an electronegative ligand. Second, and relatedly, there are sufficiently few vibrational loss channels that repumping is technologically feasible~\cite{di2004laser, Ivanov2019_rational_design,augenbraun2020molecular}. The ionic bond between the metal atom and the ligand leads to a localized orbital of the valence electron near the metal atom~\cite{Ellis2001}. Hence, when this electron is excited, it only weakly couples to bond stretching. 

Building from the above principles and approaches, it was recently shown theoretically that an alkaline earth(I)-oxide unit attached to an electron withdrawing ligand often led to a lone, metal-centered optically active electron \cite{Ivanov2020_prospects_large_organic,ivanov2020biOCC,dickerson2021franck}.  This electron’s orbital and environment is then qualitatively similar to those present in previously laser-cooled molecules and, therefore, strong optical cycling can be expected. Further, the theoretical calculations suggested that increasing the electron withdrawing strength of the ligand by substitutions in the ligand could improve optical cycling performance. Some of these concepts were recently demonstrated for ligands composed of phenyl and its derivatives  \cite{zhu2022phenol}, where it was suggested that the alkaline earth oxide unit could be considered a quantum functional group that could be used to gain control of larger molecules and provide a generic qubit moiety for attaching to large molecules and perhaps surfaces.

It is a tantalizing question whether even larger arenes would possess favorable properties for optical cycling. This idea was theoretically explored in \cite{dickerson2021optical}, which  found that larger rings in polycyclic aromatic molecules begin to close the highest occupied - lowest unoccupied molecular orbital (HOMO-LUMO) gap of the ligands relative to the electronic transition of the OCC. As the orbitals belonging to the arenes get closer in energy to the otherwise isolated electronic transition, they distort the potential energy surface and disrupt the vibrational overlap between electronic states. Therefore, as the number of rings is increased, the optical transition is expected to become progressively less diagonal. However, even in this case it may be possible to use an electron withdrawing substitution around the ring to increase Frank-Condon diagonality. It was predicted that for $\ge$10 rings (ovalene), the HOMO-LUMO gap will be closed to the point that the transition would cease to be diagonal~\cite{dickerson2021optical}. 

\begin{figure*}[ht!]
    \centering
    \includegraphics[scale=1]{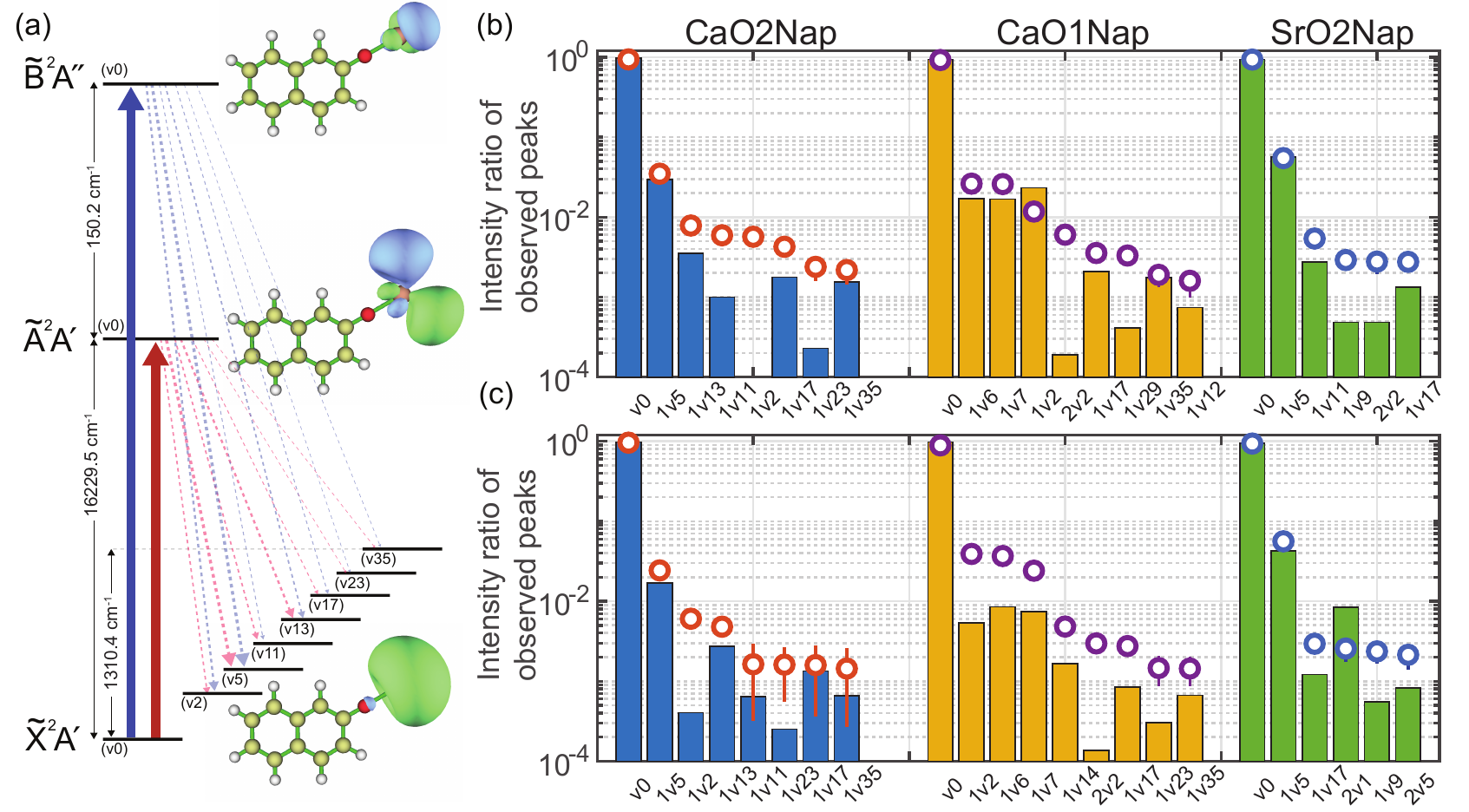}
    \caption{Energy level diagram and intensity ratios of all observed peaks for the three molecular species studied in this work and their comparison with theory. (a) Electronic energy levels (\X, \A and \B) and vibrational energy levels (v0 to v35) for CaO2Nap. Solid arrows indicate OPO excitation wavelengths. Dashed lines represent decays to vibrationally excited states (spacings are not to scale). Adjacent to the energy levels are MO representations of the ground (\X) and both excited \A (in-plane) and \B (out-of-plane) states. (b)  Intensity ratio of all observed peaks for \A $\rightarrow$ \X decay and (c) \B $\rightarrow$ \X decay. Vibrational modes are denoted as v$M$ for the $M^\text{th}$ vibrational mode and $N$v$M$ implies that $N$ vibrational quanta are excited.  Molecules are, from left to right: CaO2Nap, CaO1Nap and SrO2Nap. Circles denote experimental intensity while bars are from theory. Error bars are statistical standard errors from fits.}
    \label{fig:fig2}
\end{figure*}

Here, we take the first step towards understanding the functionalization of large polycyclic arenes with an optical cycling center, which can act as a quantum functional group. We report on the production and spectroscopic study of CaO and SrO substituted naphthyl (Nap), measuring the optical decay to vibrational states of the ground electronic state. We present a solid-precursor-based production technique, which could be generalized to produce a large variety of arenes in a CBGB. Using this molecule source and the dispersed laser-induced fluorescence (DLIF) technique~\cite{Reilly2006, Gascooke2011, Kokkin2014,augenbraun2021}, we detect and characterize the optical excitation and decay of naphthol based molecules. We measure vibrational branching ratios (VBRs), which quantify the probability of decay from an excited electronic state to each vibrational state of the ground electronic manifold, and serve as a measure of the ability to scatter a large number of photons. We observe that the molecule CaO\twonaphthyl~has a VBR for the vibrationless transition of 96(1)\%. We also compare the isomers CaO\twonaphthyl~and CaO\onenaphthyl~and observe that for the latter, the transition becomes less diagonal. Finally, we also perform spectroscopy on SrO\twonaphthyl~and find it also to be favorable for optical cycling.

\begin{figure}[t!]
    \centering
    \includegraphics[scale=1]{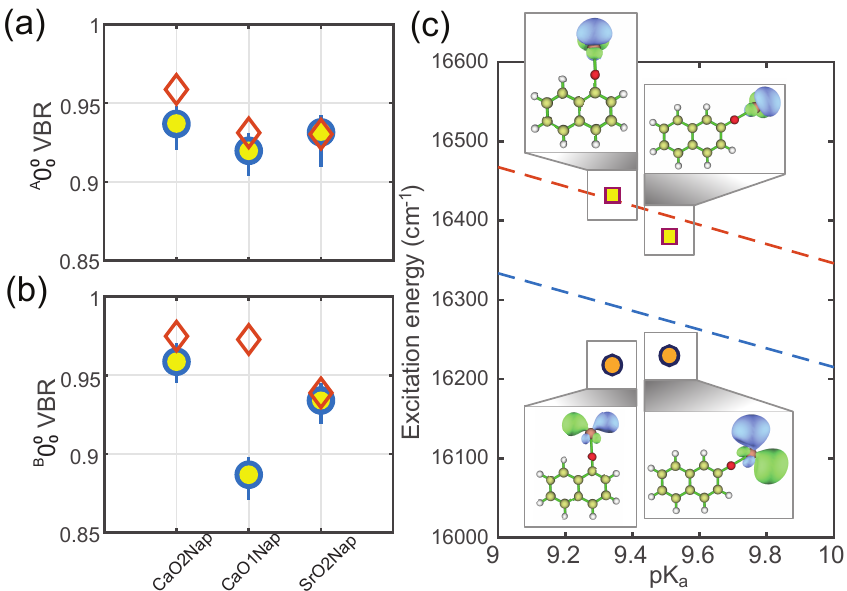}
    \caption{$0^0_0$ VBR for the three molecules studied in this work and excitation energy as a function of pK$_a$.   (a) primary VBR for the \A state ($^A0^0_0$) and (b) primary VBR for the \B state ($^B0^0_0$) for the three molecules, from left to right: CaO2Nap, CaO1Nap and SrO2Nap. Experimental results (circles) is shown together with theory (diamonds). Error bars are a combination of statistical and systematic uncertainties. (c) \X$\rightarrow$ \A excitation energies depicted as circles and \X$\rightarrow$ \B excitation energies depicted as squares for CaO1Nap (pK$_a$ = 9.34) and CaO2Nap (pK$_a$ = 9.51). Dashed line is excitation energy vs pK$_a$ trend for CaOPh-X derivatives taken from \cite{zhu2022phenol}. Insets are MO representation of each electronic state. \B state energies lie close to this trend line while \A state energies deviate significantly, likely due to in-plane orbitals in the \A state.}
    \label{fig:fig3}
\end{figure}

Details of the experimental setup are provided in the supplement \cite{Supp}. Briefly, molecules are produced in a CBGB source via ablation of a solid precursor consisting of CaH$_2$ or SrH$_2$ powder mixed with 1-naphthol or 2-naphthol powder. The molecules are cooled by the buffer gas and are optically excited with a pulsed laser (OPO). The dispersed fluorescence spectrum is obtained by collecting molecular fluorescence light through a spectrometer onto a camera.

The four lowest energy states of the molecules studied in this work can be represented as \X$^2A'$, \A$^2A'$, \B$^2A''$ and \C$^2A'$ in the $C_s$ molecular symmetry group \cite{augenbraun2020molecular}. We explored the transitions between the ground vibrational state of the ground electronic state (\X) and the ground vibrational state of the two lowest electronically excited manifolds (\A and \B, Fig. \ref{fig:fig2}(a)). In case of the \A $\rightarrow$ \X decay, we observe that the decay is predominantly to the ground-vibrational mode (denoted by $^A0^0_0$), with a small amount to the first Ca-O stretching mode $^A5^0_1$, for both CaO2Nap (Fig. \ref{fig:fig1}(b(i))) and SrO2Nap (Fig. \ref{fig:fig1}(b(iii))), and $^A6^0_1$ and $^A7^0_1$ for CaO1Nap (Fig. \ref{fig:fig1}(b(ii))). 

In the case of \B $\rightarrow$ \X decay, we observe similar behavior with the dominant decay to the ground vibrational mode ($^B0^0_0$). We also note that in every case, the excited molecules in the \B state can decay to the lower energy \A state via collisional relaxation. This process has been observed in similar experiments \cite{Liu2019collision,dagdigian1997collision,zhu2022phenol}. If a significant fraction of the molecules in the \B state decay via collisional relaxation, we expect to see not only the $^A0^0_0$ decay but also the higher $^A$v$^0_1$ vibrational decay peaks. Such collisional relaxation peaks are denoted with $*$ in Fig. \ref{fig:fig1} (c(i)-(iii)). It is interesting to note that the reverse process of collisional excitation from \A $\rightarrow$ \B is also observed in the experiment, albeit at a much lower level than the \B $\rightarrow$ \A process. This is possibly due to energetic collision partners.

We fit the DLIF spectra to a series of Pearson distributions \cite{Supp} and plot the result of the fitting in Fig. \ref{fig:fig2}(b) and (c). In principle, a measurement of the VBRs requires knowledge of all vibrational modes, including any that may not be observed either due to insufficient measurement sensitivity, obstruction by spurious peaks, or because the corresponding decays lie outside of the detected range of wavelengths. Instead, we plot the ratio of the intensity of each observed peak to the sum of the intensities of all observed peaks. We plot this quantity for both experiment and density functional theory calculations  for comparison (details can be found in the supplement \cite{Supp}). The \A $\rightarrow$ \X (Fig. \ref{fig:fig2}(b)) and \B $\rightarrow$ \X (Fig. \ref{fig:fig2}(c)) decays show good agreement with theory for the dominant decay channels ($0^0_0$ for all three, $5^0_1$ for MO2Nap, and $6^0_1$ and $7^0_1$ for CaO1Nap). Theory, however, fails to predict significant decay to the bending mode of the M-O-C bond ($2^0_1$) for MO2Nap. We speculate that this discrepancy is due to vibronic couplings among and anharmonicities within the low-frequency modes \cite{domcke1977theoFCF,Fischer1984, dickerson2021franck,zhu2022phenol}, which are beyond the scope of the calculations presented in this work. However, our calculations predict a significant bending contribution for CaO1Nap~for both the first ($2^0_1$) and second ($2^0_2$) harmonics (Fig. \ref{fig:fig2}). The contribution of the bending mode for CaO1Nap is by far the highest among all species examined here and the CaO\phenylX{X} molecules, examined in \cite{zhu2022phenol}. This is likely related to the fact that the CaO group in CaO1Nap is bonded to the carbon at position 1 of naphthalene (Fig. \ref{fig:fig1}(a(ii))). This bonding site leads to a higher proximity to nearby carbon and hydrogen atoms and hence a greater mixing with the naphthol orbitals upon bending motion \cite{Radkowska2021naphthol}.

In order to estimate the VBRs, we consider four systematic effects, as described in the supplement \cite{Supp}. These include contributions of unobserved decays, population in vibrationally excited levels of \X, fluctuations in excitation light power, and calibration of the spectrometer response. These sources of systematic uncertainty are added in quadrature with the statistical uncertainty of the fits to give upper and lower uncertainty bounds for each vibronic decay in Table S1. We show the determined diagonal VBR for the \A$\rightarrow$ \X decay in Fig. \ref{fig:fig3}(a) and the \B$\rightarrow$ \X decay in Fig. \ref{fig:fig3}(b), together with the theoretical predictions.

Our data elucidate the correlation between certain physical properties of calcium-based molecules with the electron-withdrawing property of the ligand. Naphthol can be compared with other proton donors based on their acid dissociation constant or pK$_a$. A smaller pK$_a$ implies a more ionic O-H bond. We plot the excitation energies obtained for the \A$\leftarrow$ \X and \B$\leftarrow$ \X transitions for CaO1Nap and CaO2Nap as a function of pK$_a$ of the corresponding naphthols in Fig. \ref{fig:fig3}(c) in comparison to the excitation energy trend for CaO\phenyl~and derivatives obtained from \cite{zhu2022phenol}. The \B$\leftarrow$ \X excitation energies closely follow the trend defined by the \B$\leftarrow$ \X transitions of the phenol derivatives. However, the \A$\leftarrow$ \X excitation energies deviate significantly from the trend. We attribute this effect to the fact that the \B state orbital is out-of-plane to the naphthalene rings while the \A state orbital is in-plane \cite{Morbi1997c2v}. Due to the proximity with other carbon and hydrogen atoms, the bending force constant of the \A state can significantly deviate from a pure CaO\phenylX{X} bond (see Fig. \ref{fig:fig3}(c) insets).  

\begin{table}[htb]
\centering
\begin{tabular}{|c|c|c|c|}
\hline
Molecule                & State & Excitation wavelength (nm) & Lifetime (ns) \\ \hline
\multirow{2}{*}{CaO\twonaphthyl} & \A     & 616.16(5)                   & 19.8(2.5)         \\ \cline{2-4} 
                        & \B     & 610.51(5)                   & 24.4(1.8)         \\ \hline
\multirow{2}{*}{CaO\onenaphthyl} & \A     & 616.61(5)                   & 27.4(2.5)         \\ \cline{2-4} 
                        & \B     & 608.56(5)                   & 22.4(2.4)         \\ \hline
\multirow{2}{*}{SrO\twonaphthyl} & \A     & 668.02(5)                   & 28.5(2.8)         \\ \cline{2-4} 
                        & \B     & 654.43(5)                   & 26.7(2.9)         \\ \hline
\end{tabular}
\caption{Excitation wavelength relative to the ground vibrational (\X) state (in nm) and measured lifetime (in ns) for the non-vibrating \A and \B electronic excited states of the three molecules measured in this work. The uncertainty in wavelength estimation is the same for all measurements and is given by the spectral resolution ($\sim$ 0.05~nm) of the optical spectrum analyzer used to measure the OPO center wavelength, which in turn was used to convert the camera pixels to wavelength. The lifetimes are obtained from exponential fits to the DLIF signals as a function of delay. The error bars are standard errors of fits.}
\label{tab:table1}
\end{table}

Finally, in order to understand the potential for optical cycling and laser cooling, which is determined not only by the VBR, but also by the strength of the optical transition, we measure the lifetime of the \A and \B states. This is done by varying the delay between the OPO excitation and camera acquisition in steps of 5~ns (see Table \ref{tab:table1}). The lifetimes for all measured states are roughly 20 to 30~ns, similar to other molecules belonging to the CaOX family and compatible with maximum photon scattering rates of $\sim10^6$ s$^{-1}$~\cite{baum20201d}. 

In conclusion, we demonstrate that a class of molecules based on a bicyclic aromatic compound, naphthalene, is favorable for optical cycling when functionalized with a calcium- or strontium- based optical cycling center. We establish a method that allows for the chemical formation of naphthol derivatives in a CBGB and could possibly be extended to larger arenes. Using DLIF spectroscopy, we measure that $\approx$96\% of photon scattering events do not result in a change in the vibrational state of the molecule CaO\twonaphthyl, which should allow for efficient optical cycling. We elucidate the role of geometry in decoupling the electronic and vibrational degrees of freedom by showing that the spatial proximity of the ligand in CaO\onenaphthyl~causes a measurable decrease in $0^0_0$ VBR and stronger excitation of other vibrational modes. The use of Sr (instead of Ca) might provide a technological advantage through the use of laser-diode-accessible transitions, while still providing a high $0^0_0$ VBR ($\approx$93\%). Rotational closure for asymmetric top molecules is possible with 1-2 lasers per vibrational repump \cite{augenbraun2020molecular}, enabling rapid photon cycling. Optically enabled quantum state control of aromatic molecules could enable studies of chemical reactions via molecular collisions \cite{Augustoviov2019collisions,Cheuk2020collisions}. This work provides the stepping stone for the study of even larger aromatic compounds \cite{dickerson2021optical} or even surfaces \cite{guo2021surface} with quantum control over all degrees of freedom, opening new avenues in quantum chemistry.  Finally, these measurements indicate that rapid photon cycling will be possible, which will enable laser cooling to the ultracold regime and fast and efficient quantum state manipulation of large arene molecules.      

The authors thank Timothy C. Steimle for sharing critical equipment and for useful discussions.  MO visualizations were performed using Multiwfn \cite{Lu2012miltiwfn}.
This work was supported by grants from the Keck Foundation, the CUA, the DOE, National Science
Foundation (Grants No. PHY-1255526, No. PHY-1415560, No. PHY-1912555, No. CHE-1900555,
and No. DGE-1650604) and Army Research Office (Grants
No. W911NF-15-1-0121, No. W911NF-14-1-0378,
No. W911NF-13-1-0213 and  W911NF-17-1-0071) grants. C.E.D. would like to acknowledge NSF GFRP grant No. DGE-2034835.

\providecommand{\latin}[1]{#1}
\makeatletter
\providecommand{\doi}
  {\begingroup\let\do\@makeother\dospecials
  \catcode`\{=1 \catcode`\}=2 \doi@aux}
\providecommand{\doi@aux}[1]{\endgroup\texttt{#1}}
\makeatother
\providecommand*\mcitethebibliography{\thebibliography}
\csname @ifundefined\endcsname{endmcitethebibliography}
  {\let\endmcitethebibliography\endthebibliography}{}

\pagebreak

\clearpage

\setcounter{equation}{0}
\setcounter{table}{0}
\setcounter{figure}{0}
\renewcommand{\thefigure}{S\arabic{figure}}
\renewcommand{\theequation}{S\arabic{equation}}
\renewcommand{\thetable}{S\arabic{table}}

\renewcommand{\thefigure}{S\arabic{figure}}

\renewcommand{\theequation}{S\arabic{equation}}

\centerline{\Large\textbf{Supplemental Information}}

\bigskip

\normalsize
\section{Experimental setup}
The experimental setup consists of a cryogenic buffer-gas beam (CBGB) source operating at $\approx$ 9~K \cite{hutzler2012buffer}. The solid precursor target (described below) is ablated with the second harmonic of a Nd:YAG pulsed laser at 532~nm (20~mJ per pulse, $\sim$ 5~ns pulse duration and 10~Hz repetition rate). The resulting ablation effluent is cooled via collisions with helium buffer gas at densities of $\approx 10^{15}$~cm$^{-3}$  that is flowed into the cell through a capillary. This gas-phase molecular mixture includes the molecule of interest. Because the temperature of the buffer gas is much smaller than the vibrational splittings ($\approx$100~K) and on the order of the rotational splittings ($\approx$1~K), only a handful of low rovibrational states are populated. In order to optically excite the molecule of interest, we employ a pulsed optical parametric oscillator (OPO), which can be continuously tuned from 500 to 700 nm. The excitation pulse is 10 ns long and has a spectral width of $\sim$ 0.2 nm. The fluorescence photons are collected with a 35 mm focal length lens placed on the cell and guided with mirrors into a 0.67~m focal length Czerny-Turner style monochromator. The dispersed fluorescence is imaged onto a gated, intensified charge-coupled device
camera (ICCD) cooled to $-30$~$^\circ$C. 
A roughly $80$~nm wide spectral region of the DLIF can be recorded in a single camera image.

The gas-phase chemical reaction method utilized in earlier work to produce CaOPh-X \cite{zhu2022phenol} is difficult for larger arenes because of the high melting point of the reactants. For example, 2-hydroxynaphthalene (2-naphthol) has a melting point of 120$^\text{o}$C compared to 40$^\text{o}$C for hydroxybenzene (phenol). Such high temperatures are a challenge to achieve in the CBGB apparatus. Instead we prepare ablation targets by combining CaH$_2$ or SrH$_2$ powder (99.9\% trace metal basis, Sigma-Aldrich) with 1-naphthol or 2-naphthol powder (99\% purity, Sigma-Aldrich) in a 1:2 molar ratio. The mixture is finely ground and compressed into pellets at a pressure of $\approx$ 5,000~psi. We verify the chemical composition of the pellets by performing $^{13}$C NMR spectroscopy and mass spectrometry. Our studies indicate that naphtholate salts of calcium are not formed at room temperature, but the mixed pellets do allow us to vaporize both component species simultaneously through ablation. The reaction between the two components occurs within the hot ablation plume inside of the CBGB.

\section{Data acquisition}
Data are obtained in three complementary ways. First, low-resolution ($\sim$200 averages) scans are performed while varying the OPO excitation wavelength to locate the molecular transitions. These scans have the additional advantage of revealing impurities in the spectra of interest due to the presence of other molecular species that need to be omitted from calculations of VBRs. The second way is to fix the OPO wavelength to be resonant with a molecular transition and collect fluorescence photons over thousands of ablation shots in order to measure VBRs. Representative DLIF spectra obtained this way provide $\sim10^{-3}$ precision in VBR and are depicted in Fig. 1 of the main text. Third, we vary the delay between the OPO excitation pulse and the intensification stage of the ICCD camera. As this delay is increased, we measure the exponential decay in population from the excited state due to the natural radiative lifetime of the state. An exponential fit to the decay in population gives a reliable measure of the excited state lifetime.

\section{Fitting procedure}
In order to extract the intensities of the different peaks and obtain the vibrational branching ratio, we employ a fitting protocol described in detail in \cite{zhu2022phenol}. Briefly, we fit the entire spectrum to a series of Pearson distributions. Each peak is fit to two Pearson distributions, one describing the center of the peak and the other describing the tails. The shape parameters are simultaneously fit for each peak in the spectrum and only their locations and amplitudes are fit independently, thus ensuring that the shape of each peak is the same and representative of the instrument response function. We occasionally observe peaks that cannot be attributed to any theoretically calculated vibrational modes, their overtones or hybrid modes (marked with $\dagger$ in Fig. 1 of the main text). We fit these peaks and those induced by electronic-state-changing collisions (marked with $*$) but exclude them during VBR estimation. Note also that the largest rotational constant for this class of molecules is expected to be $<1$~\cm. Hence the excitation OPO light (spectral width of $\sim$ 5~\cm) could excite some rotationally excited states that remain populated after buffer gas cooling. However, the VBRs extracted from the data represent the purely vibrational branching ratios since the spectrometer (resolution $\sim$ 5~\cm) likewise cannot resolve the rotational spectrum of emitted photons.    

\section{Systematic uncertainty estimation}
The four systematic effects for the vibrational branching ratios considered in this work are described below. First, to find the probability that a molecule undergoes a particular vibronic decay, it is necessary to estimate the probability that a molecule undergoes a vibronic transition that was unobserved or unresolved. We attempt to estimate this probability by using theoretically predicted branching fractions to unobserved states, adjusted by an empirical factor that accounts for the tendency of theoretical predictions to underestimate off-diagonal decay probabilities. Specifically, we compute the average factor, $C$, by which theoretical off-diagonal branching ratios underestimate observed off-diagonal branching ratios:

\begin{equation}
C=\frac{1}{N_{\rm{od}}}\sum_{i=1}^{N_{\rm{od}}}\frac{T_i}{S_i},
\end{equation}

\noindent where $N_{\rm{od}}$ is the number of off-diagonal peaks observed, $T_i$ is the theoretically predicted probability of decaying to the $i^{\rm{th}}$ observed excited vibrational state (conditional on decaying to an observed peak) and $S_i$ is the fraction of observed decays that populate the $i^{\rm{th}}$ observed excited vibrational state. Typically, we find $C\approx0.5$, indicating that predictions underestimate off-diagonal decays by approximately a factor of 2. We then compute the theoretically predicted probability, $p_{\rm{unobs}}^{(\rm{theory})}$ of decaying to any unobserved peak, and estimate the actual probability of decaying to any unobserved peak as $p_{\rm{unobs}}^{\rm{(est)}}=p_{\rm{unobs}}^{(\rm{theory})}/C$. With this method, we estimate that (depending on the molecule and state) less than 0.5$-$2\% of decays populate unobserved vibrational states. We include the possibility of decays to unobserved states with probability $p_{\rm{unobs}}^{\rm{(est)}}$ as a source of systematic uncertainty in our determination of VBRs.

As a second source of systematic uncertainty, vibrationally excited states in the \X manifold could be near-resonantly excited by the pulsed OPO to the corresponding vibrationally excited states in the \A or \B manifolds. These vibrationally excited states would display a slightly different relative prominence of diagonal vs. off-diagonal decays, compared with \A$(\text{v}=0)$ and \B$(\text{v}=0)$, and can therefore contribute a systematic error in vibrational branching ratio estimates. Because these vibrationally excited states would exhibit blue-shifted DLIF peaks, we can limit their contribution to the diagonal vibrational branching ratio at $\pm0.5\%$. Third, fluctuations in the pulsed OPO power could cause imperfect background subtraction at the level of $\pm0.5\%$ of the dominant decay peak. Fourth, we estimate a $\pm1\%$ uncertainty in the calibration of the wavelength-dependent response of the spectrometer \footnote{Details can be found in the supplementary information section in \cite{zhu2022phenol}}.

\section{Theoretical calculations}
We perform theoretical calculations in order to identify the molecular vibrational modes and predict branching ratios. Theoretical geometry optimizations and frequencies are calculated using density functional theory (DFT)/time-dependent DFT at the PBE0-D3/def2-TZVPPD level of theory on a superfine grid with Gaussian16 \cite{perdew1996rationale,grimme2010consistent,rappoport2010property,frisch2016gaussian}. Molecular orbitals were generated with an isosurface of 0.03 in Multiwfn \cite{Lu2012miltiwfn}. Franck Condon factors (FCFs), which are the square of the overlap integral of the vibrational wavefunctions, are calculated using Duschinsky rotations with ezFCF \cite{gomez2021ezspectra}. The FCFs are subsequently converted to VBRs using the harmonic frequency of the mode \cite{Kozyryev2019FCF}.

For CaO2Nap and SrO2Nap, the \B state lies close in energy to the \A state, and as noted in an earlier paper, they could perhaps cross \cite{dickerson2021optical}. To further investigate this feature, we displace SrO2Nap and CaO2Nap geometries along the 1D v1/v2 vibrational modes to create 1D potential energy surfaces.  We find the SrO2Nap potential energy surfaces along these modes are quite anharmonic and nearly degenerate. This could explain the 2$^\text{nd}$ harmonic decays that are observed for SrO2Nap but not for CaO2Nap (Fig. 2c of the main text).  Additionally, we find the orbital pictures show evidence of \A~--~\B state crossing at large displacements, suggesting multireference methods are needed for correct nonadiabatic couplings/state crossings. However, we note that DFT theory predictions in the adiabatic picture do a suitable qualitative job to describe these 2$^\text{nd}$ harmonic decays when compared with experiment. We note that more sophisticated techniques have been developed recently to accurately predict VBRs below $10^{-4}$ in polyatomic molecules such as CaOH and YbOH \cite{Zhang2021accurateVBR}. 

\begin{table*}[htbp]
    \centering
    \renewcommand{\arraystretch}{1.2}
    \begin{tabular}{c c c c c c c}
    \hline
    \hline
    \multicolumn{7}{c}{CaO2Nap} \\
    \cline{1-7}
    Modes & Freq. (Exp. \cm)  & Freq. (Theo. \cm) & VBR (Exp. \A) & VBR (Theo. \A) & VBR (Exp. \B) & VBR (Theo. \B)  \\
    \cline{1-7}
v0 & 0.0(0.4) & 0.0 & $0.9368^{0.0107}_{0.0155}$ & 0.9587 & $0.9586^{0.0113}_{0.0133}$ & 0.9748 \\
1v2 & 56.3(3.5) & 48.8 & $0.0057^{0.0012}_{0.0012}$ & 0.0000 & $0.0060^{0.0017}_{0.0017}$ & 0.0004 \\
1v5 & 276.5(0.9) & 275.9 & $0.0349^{0.0009}_{0.0010}$ & 0.0291 & $0.0243^{0.0013}_{0.0013}$ & 0.0167 \\
1v11 & 525.7(2.9) & 530.4 & $0.0059^{0.0009}_{0.0009}$ & 0.0010 & $0.0016^{0.0013}_{0.0013}$ & 0.0006 \\
1v13 & 558.3(2.3) & 557.3 & $0.0079^{0.0009}_{0.0009}$ & 0.0035 & $0.0048^{0.0013}_{0.0013}$ & 0.0027 \\
1v17 & 784.6(3.7) & 787.4 & $0.0043^{0.0008}_{0.0008}$ & 0.0018 & $0.0016^{0.0012}_{0.0012}$ & 0.0013 \\
1v23 & 919.1(11.9) & 938.0 & $0.0024^{0.0008}_{0.0008}$ & 0.0002 & $0.0016^{0.0011}_{0.0011}$ & 0.0003 \\
1v35 & 1310.4(6.3) & 1331.2 & $0.0022^{0.0008}_{0.0008}$ & 0.0015 & $0.0014^{0.0012}_{0.0012}$ & 0.0007 \\
    \hline
    \multicolumn{7}{c}{CaO1Nap}  \\
    \cline{1-7}
    Modes & Freq. (Exp. \cm)  & Freq. (Theo. \cm) & VBR (Exp. \A) & VBR (Theo. \A) & VBR (Exp. \B) & VBR (Theo. \B)  \\
    \cline{1-7}
v0 & 0.0(0.2) & 0.0 & $0.9198^{0.0106}_{0.0149}$ & 0.9312 & $0.8868^{0.0103}_{0.0140}$ & 0.9727 \\
1v2 & 47.6(1.8) & 50.6 & $0.0117^{0.0015}_{0.0015}$ & 0.0231 & $0.0390^{0.0013}_{0.0014}$ & 0.0054 \\
2v2 & 103.3(3.1) & 101.2 & $0.0060^{0.0008}_{0.0008}$ & 0.0002 & $0.0030^{0.0008}_{0.0008}$ & 0.0001 \\
1v6 & 268.7(1.4) & 282.2 & $0.0262^{0.0007}_{0.0008}$ & 0.0169 & $0.0367^{0.0009}_{0.0010}$ & 0.0085 \\
1v7 & 324.1(0.7) & 323.4 & $0.0259^{0.0007}_{0.0007}$ & 0.0167 & $0.0241^{0.0007}_{0.0007}$ & 0.0074 \\
1v12 & 538.1(6.8) & 558.8 & $0.0016^{0.0006}_{0.0006}$ & 0.0007 & - & 0.0003 \\
1v14 & 636.9(2.3) & 639.5 & - & 0.0011 & $0.0048^{0.0006}_{0.0006}$ & 0.0017 \\
1v17 & 761.1(4.1) & 769.7 & $0.0036^{0.0006}_{0.0006}$ & 0.0021 & $0.0028^{0.0006}_{0.0006}$ & 0.0008 \\
1v23 & 915.3(7.4) & 915.3 & - & 0.0001 & $0.0015^{0.0006}_{0.0006}$ & 0.0003 \\
1v29 & 1074.6(3.2) & 1123.0 & $0.0033^{0.0006}_{0.0006}$ & 0.0004 & - & 0.0002 \\
1v35 & 1319.1(7.7) & 1340.2 & $0.0019^{0.0006}_{0.0006}$ & 0.0017 & $0.0014^{0.0006}_{0.0006}$ & 0.0007 \\
    \hline
    \multicolumn{7}{c}{SrO2Nap}  \\
    \cline{1-7}
    Modes & Freq. (Exp. \cm)  & Freq. (Theo. \cm) & VBR (Exp. \A) & VBR (Theo. \A) & VBR (Exp. \B) & VBR (Theo. \B)  \\
    \cline{1-7}
v0 & 0.0(0.4) & 0.0 & $0.9315^{0.0105}_{0.0202}$ & 0.9303 & $0.9340^{0.0106}_{0.0140}$ & 0.9388 \\
2v1 & 98.6(4.3) & 77.6 & - & 0.0000 & $0.0026^{0.0008}_{0.0008}$ & 0.0083 \\
2v2 & 95.2(3.8) & 87.4 & $0.0028^{0.0008}_{0.0008}$ & 0.0005 & - & 0.0003 \\
1v5 & 207.6(0.5) & 208.7 & $0.0547^{0.0009}_{0.0013}$ & 0.0563 & $0.0560^{0.0009}_{0.0011}$ & 0.0419 \\
2v5 & 410.1(5.0) & 417.4 & - & 0.0016 & $0.0021^{0.0007}_{0.0007}$ & 0.0008 \\
1v9 & 465.9(4.9) & 468.2 & $0.0029^{0.0007}_{0.0007}$ & 0.0005 & $0.0024^{0.0007}_{0.0007}$ & 0.0005 \\
1v11 & 520.6(1.9) & 523.3 & $0.0054^{0.0007}_{0.0007}$ & 0.0027 & - & 0.0025 \\
1v17 & 765.2(5.2) & 775.7 & $0.0028^{0.0006}_{0.0006}$ & 0.0013 & $0.0029^{0.0007}_{0.0007}$ & 0.0012 \\
    \hline
    \hline
    \end{tabular}
    \caption{Observed and theoretical vibrational mode frequencies and VBRs for all molecules studied in this work. Frequency error bars are standard error of the fits. Asymmetric error bars for the experimental VBRs account for the standard error of fit as well as known systematic uncertainties for the experiment.}
    \label{tab:table2}
\end{table*}

\providecommand{\latin}[1]{#1}
\makeatletter
\providecommand{\doi}
  {\begingroup\let\do\@makeother\dospecials
  \catcode`\{=1 \catcode`\}=2 \doi@aux}
\providecommand{\doi@aux}[1]{\endgroup\texttt{#1}}
\makeatother
\providecommand*\mcitethebibliography{\thebibliography}
\csname @ifundefined\endcsname{endmcitethebibliography}
  {\let\endmcitethebibliography\endthebibliography}{}

\end{document}